\documentclass[twocolumn]{openjournal}
\usepackage{amsmath,amssymb,physics,bm,esint}
\usepackage{physics,commath}
\usepackage{orcidlink}
\usepackage{color}
\usepackage{listings}
\usepackage{graphicx}
\usepackage{textcomp}
\usepackage{soul}
\usepackage{comment}
\usepackage{placeins} 
\usepackage{xlmacros}
\usepackage{tikz}
\usepackage{lineno}

\DeclareRobustCommand{\VAN}[3]{#2}
\let\VANthebibliography\thebibliography
\def\thebibliography{\DeclareRobustCommand{\VAN}[3]{##3}\VANthebibliography}

\begin{document}

\title{Analytical weak-lensing shear response of galaxy model fitting}

\author{Xiangchong Li}
\email{xli6@bnl.gov}
\affiliation{Brookhaven National Laboratory, Bldg 510, Upton, New York 11973, USA}

\submitted{Received Sep. 10, 2026; accepted Sep. 30, 2026}

\begin{abstract}
    Galaxy model fitting is widely employed to estimate properties such as
    galaxy shape, size, and color. Understanding how the outputs of galaxy
    model fitting respond to weak lensing shear distortions is crucial for
    accurate shear estimation and mitigating shear-related systematics in weak
    lensing image analyses. In this paper, we investigate how the fitted
    parameters---specifically flux, size, and shape---respond to weak-lensing
    shear distortions within the \anacal{} framework. To achieve this, we
    introduce \emph{quintuple numbers}, a novel algebraic system inspired by
    dual numbers from automatic differentiation. Quintuple numbers enable the
    propagation of shear response information throughout the entire
    model-fitting process by linking analytical pixel shear responses to those
    of the fitted parameters. We integrate \emph{quintuple numbers} into the
    \anacal{} framework to derive the shear responses of shapes estimated with
    model fitting and validate the pipeline using image simulations that
    include realistic blending. Our results demonstrate that the multiplicative
    bias remains below $3\times10^{-3}$ for ground-based, oversampled images.
\end{abstract}


\maketitle

\section{INTRODUCTION}

We are entering an exciting era of precision cosmology, driven by the launch of
next-generation astronomical imaging surveys \citep{DarkEnergy}. These
groundbreaking projects aim to rigorously test the prevailing cosmological
model and shed light on the accelerated expansion of the Universe. A key
observational tool in this endeavor is weak gravitational lensing, which traces
the growth of cosmic structure. This phenomenon arises when massive structures
bend the light from distant galaxies, introducing subtle, coherent shear
distortions in their observed shapes \citep{rev_wl_Bartelmann01,
rev_cosmicShear_Kilbinger15, rev_wlsys_Mandelbaum2017}.

Among the most prominent ``Stage IV'' surveys are the Vera C.\ Rubin
Observatory's Legacy Survey of Space and Time (LSST,
\citealt{LSSTOverviwe2019}), the European Space Agency’s Euclid mission
\xlrv{\citep{Euclid2011, EuclidMellier2025}}, and NASA’s Nancy Grace Roman Space Telescope
\xlrv{\citep{Roman2020, Spergel2015}}. To meet the science goals, these programs demand shear
measurement techniques with systematic biases controlled at the level of a few
tenths of a percent \citep{WLsys_Massey2013, LSSTRequirement2018}. Achieving
this level of precision requires survey-specific algorithmic solutions tailored
to the unique characteristics of each dataset.

Weak gravitational lensing introduces a coherent distortion to galaxy shapes
through a remapping of image coordinates. This effect can be described by a
Jacobian transformation that maps the true (source) coordinates
$\boldsymbol{\beta}$ to the observed (image) coordinates $\boldsymbol{\theta}$
via the lensing potential. In the weak lensing regime, where deflections are
small, this mapping can be approximated as a linear transformation:
\begin{equation}
\boldsymbol{\theta} = \mathbf{A} \cdot \boldsymbol{\beta},
    \quad \text{where} \quad \mathbf{A} = \begin{pmatrix}
1 - \kappa - \gamma_1 & -\gamma_2 \\
-\gamma_2 & 1 - \kappa + \gamma_1
\end{pmatrix}\,.
\end{equation}
Here, $\kappa$ is the convergence, which magnifies or demagnifies the image,
and $(\gamma_1, \gamma_2)$ are the two components of the shear distortion,
which introduce anisotropic stretching. In this paper, we assume $\kappa = 0$
and focus exclusively on the shear components $\gamma_1$ and $\gamma_2$.
\xlrv{More precisely, in the weak-lensing approximation the quantity
that can be observed is the reduced shear $g \equiv \gamma/(1-\kappa)$, which
reduces to $\gamma$ in the limit $\kappa \to 0$ adopted here.} In
practice, weak lensing measurements rely on statistically extracting the
\xlrv{reduced shear $\boldsymbol{g}$} from the observed ellipticity of many galaxies, under the
assumption that their intrinsic shapes are randomly oriented. A key to accurate
shear estimation lies in computing the linear shear response of galaxy shape
measurements---that is, the first-order derivative of the measured shape with
respect to the applied shear \citep{metacal_Huff2017}. By evaluating this
response, one can construct a perturbative solution that relates observed shape
distortions to the underlying shear field, enabling shear inference with
accuracy at the sub-percent level, even in the presence of complex image
systematics.

In recent years, several innovative approaches to weak lensing shear estimation
have emerged that reduce or eliminate the need for external image
simulations---particularly in the case where blended sources reside at the same
redshift. Among these, \metadet{} \citep{metaDet_Sheldon2020, metaDet_LSST2023}
stands out for its use of numerical self-calibration, while \anacal{}
\citep{Anacal_Li2023, Anacal_Li2025} introduces a fully analytical method for
deriving the linear shear response. Thanks to its analytical nature, \anacal{}
is exceptionally efficient: it can compute a shear estimator for a single
detected galaxy in under one millisecond. Another promising approach is \BFD{}
\citep{BFD_Bernstein2014, BFD_Bernstein2016}, a Bayesian framework for shear
estimation. However, \BFD{} currently may require percent-level corrections
from simulations to mitigate detection biases, particularly in cases involving
galaxy blending. \xlrv{A complementary strategy is to forward-model
the shear effects while simultaneously minimizing the sensitivity to image
simulations, which is the default approach adopted for Euclid
\citep{lensfit2024}.} It is worth noting that external simulations may still be
necessary to calibrate the redshift distribution of source galaxies when
blending occurs between galaxies at different redshifts, each subject to
distinct shear distortions \citep{DESY3_BlendshearCalib_MacCrann2021}. In this
paper, we concentrate on correcting shear estimation bias assuming blended
sources are from the same redshift, and defer the complexities introduced by
multi-redshift blending to future work.

The analytical formalism introduced by \citet{Anacal_Li2023, Anacal_Li2024}
provides a unified approach for correcting four major sources of shear bias:
detection bias, selection bias, galaxy model bias and Point Spread Function
(PSF) shape leakage. It achieves accuracy up to second order in shear. The
method begins by computing the shear response of each smoothed image pixel
value, then propagates these responses to downstream quantities---such as
detection likelihood, selection metrics, and shape estimators---using the chain
rule of derivatives. This pixel-level formulation allows for a rigorous
analytical connection between the underlying shear field and the final shear
estimator. Building on this foundation, \citet{Anacal_Li2025} extend the
framework to include a correction for noise bias. This is accomplished by
generating synthetic, pure-noise images that replicate the noise
characteristics of the observed data, akin to the approach used in
metacalibration \citep{metacal_Sheldon2017}. These noise realizations allow the
analytical formalism to robustly account for biases introduced by stochastic
image noise.

In this paper, we extend the \anacal{} formalism to derive the shear response
of the galaxy model-fitting process. This advancement enables a quantitative
understanding of how fitted model parameters---such as galaxy flux, shape, and
size---respond to weak lensing shear distortions. To accomplish this, we
introduce quintuple numbers, a novel commutative ring structure that allows the
automatic propagation of shear responses from the smoothed image pixels (inputs
to model fitting) to the model parameters (outputs of the fitting). The
quintuple number system is inspired by dual numbers, which are commonly used in
automatic differentiation to track first-order derivatives during computation
\citep{autodiff}. In analogy, quintuple numbers are designed to carry
first-order shear response terms throughout the entire model-fitting process.
This allows for a fully analytical and efficient computation of how each model
parameter varies with shear, without relying on numerical perturbations or
external simulations.

This paper is organized as follows: Section~\ref{sec:method_pixres} introduces
the analytical shear response of model fitting process and develops a shear
estimator with Gaussian model fitting. Section~\ref{sec:sim} presents
systematic tests on ground-based image simulations with the observational
condition of Hyper Suprime-Cam Survey \citep{HSC3_catalog}. Finally,
Section~\ref{sec:summary} provides a summary and outlook.

\section{SHEAR RESPONSE OF MODEL FITTING}
\label{sec:method_pixres}

In this section, we provide a comprehensive overview of our methodology. We
begin in Section~\ref{subsec:method_pixres} by reviewing the analytical pixel
shear response introduced in \citet{Anacal_Li2023} and noise bias correction
introduced in \citet{Anacal_Li2025}. In Section~\ref{subsec:method_qnumber}, we
introduce the quintuple number system that is utilized to propagate our
analytical pixel shear response. Finally, in Section~\ref{subsec:method_fit},
we describe a simple model-fitting procedure used for shape estimation, and use
quintuple numbers to derive shear response of its output parameters.

\subsection{PSF and Noise Correction}
\label{subsec:method_psfnoise}

\begin{figure*}
\centering
\includegraphics[width=.92\textwidth]{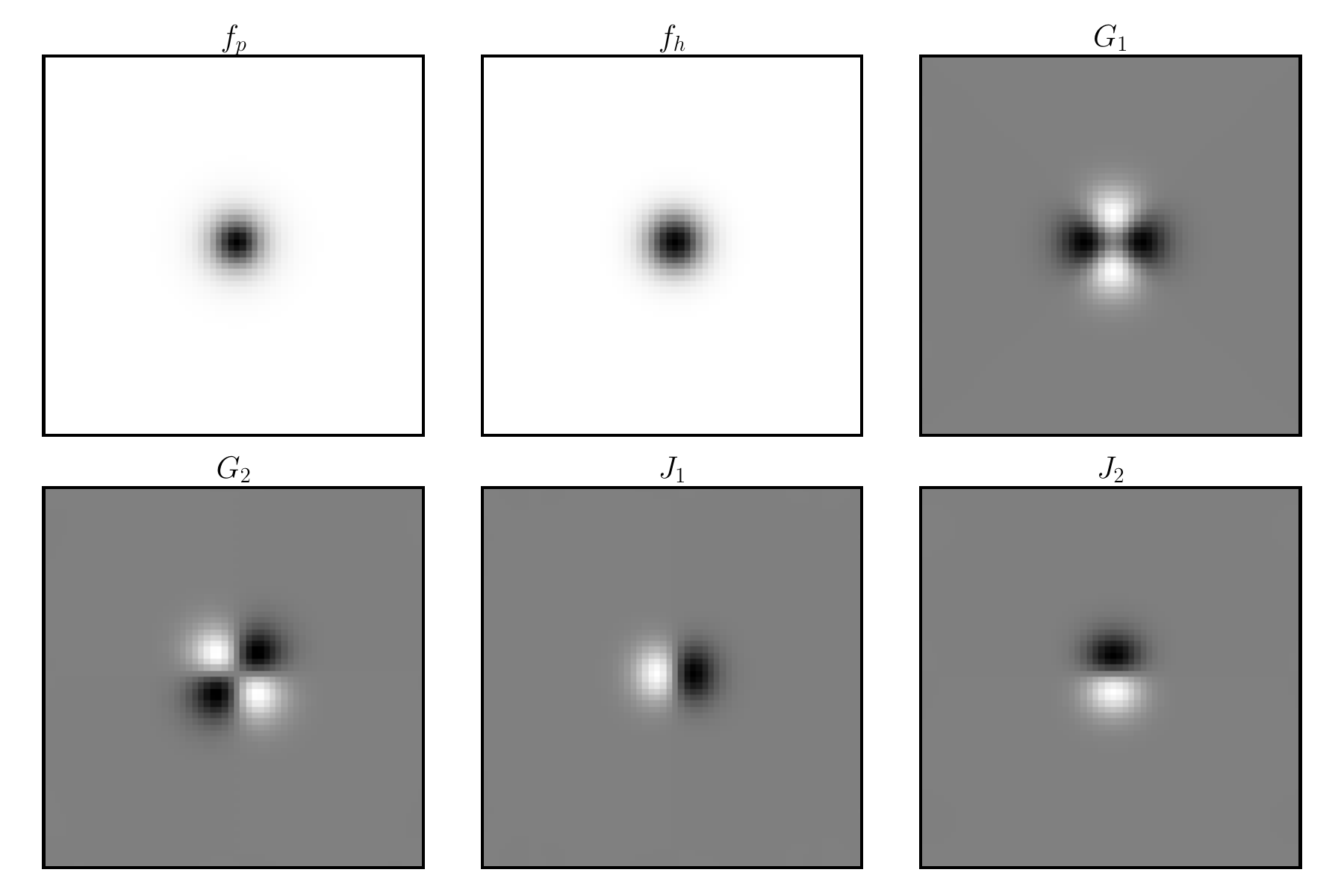}
\caption{
    We show the noiseless image $f_p(\mathbf{x})$, the smoothed image
    $f_h(\mathbf{x})$, and the filtered images $(G_1(\mathbf{x})\,,
    G_2(\mathbf{x})\,, J_1(\mathbf{x})\,, J_2(\mathbf{x}))$, which are key
    components of the pixel-level shear response \citep{Anacal_Li2023}. The
    images $G_1$ and $G_2$ represent the response of the smoothed image
    $f_h(\mathbf{x}) $ to small shear distortions along the $\gamma_1$ and
    $\gamma_2$ directions, respectively. In contrast, $J_1$ and $J_2$ quantify
    how this shear response changes with shifts in the reference point about
    which the shear distortion is applied. The original image has a seeing size
    (FWHM) of $0\farcs80$, and the FWHM of the Gaussian smoothing kernel is
    $0\farcs85$.
}\label{fig:pixel_res}
\end{figure*}

In this study, we define the observed image smeared by the PSF as $f_p(\vx)$,
where $p(\vx)$ denotes the PSF model. We assume that the PSF model is both
accurate and precise, and we refer readers to \citep{2023MNRAS.525.2441Z,
2025arXiv250500093B} for PSF modelling errors. The Fourier transforms of the
observed image and the PSF model are defined as
\begin{equation}
\begin{split}
    f_p(\vk) &= \iint\! \rmd^2 x\, f_p(\vx) \exp(-\rmi\, \vk \cdot \vx)\,,\\
    p(\vk) &= \iint\! \rmd^2 x\, p(\vx) \exp(-\rmi\, \vk \cdot \vx)\,.
\end{split}
\end{equation}

To eliminate the anisotropy in the original PSF and enable the use of the
analytical pixel shear response method proposed by \citet{Anacal_Li2023}, we
transform the PSF into an isotropic Gaussian form. This is achieved by first
deconvolving the original PSF and then applying an isotropic Gaussian filter in
Fourier space (see \citealt{Z08, metacal_Huff2017, metacal_Sheldon2017} for
similar procedures). The Gaussian filter, with standard deviation $\sigma_h$,
is defined as
\begin{equation}
    h(\vk) = \exp\!\left(-\frac{k^2 \sigma_h^2}{2} \right)\,,
\end{equation}
where $k \equiv \abs{\vk}$ is the magnitude of the wave-number vector. As will
be shown in Section~\ref{subsec:method_pixres}, this procedure facilitates the
derivation of shear response from the smoothed image.

To remove noise bias in shear estimation, we generate a pure noise image with a
different realization but identical statistical properties to the noise in the
observed image. This pure noise is then rotated by $90^\circ$ and added to the
observed image to create a renoised version, following the approach of
\citet{metacal_Sheldon2017}. We denote the Fourier transform of this pure noise
as $n(\vk)$, and its deconvolution uses the PSF rotated by $90^\circ$, denoted
$p^{90}(\vk)$. We then apply the analytical noise bias correction introduced in
\citet{Anacal_Li2025}. Specifically, the noise is deconvolved using
$p^{90}(\vk)$ and then convolved with the Gaussian $h(\vk)$. The resulting
renoised, smoothed image is denoted as $f_h(\vx)$, and its Fourier transform is
given by
\begin{equation}
\label{eq:quintuple_image_value}
f_{h}(\vk) =  h(\vk) \left(
    \frac{f_{p}(\vk)}{p(\vk)} + \frac{n(\vk)}{p^{90}(\vk)}
\right) \,.
\end{equation}

\xlrv{To make this sequence of operations explicit, the smoothed,
renoised image of equation~\eqref{eq:quintuple_image_value} is constructed
entirely in Fourier space through the following steps. (i)~The observed image is
\emph{deconvolved} by dividing its Fourier transform by the PSF model $p(\vk)$,
which removes the (generally anisotropic) PSF. (ii)~It is then \emph{reconvolved}
with the isotropic Gaussian kernel $h(\vk)$; because $h$ is round, the effective
PSF of the resulting image is isotropic, which is precisely the condition under
which the analytical pixel shear response of \citet{Anacal_Li2023} holds. The
Gaussian scale $\sigma_h$ is set so that the reconvolved profile is
broader than the original PSF: this regularizes the deconvolution and prevents
the amplification of small-scale noise at high $k$, where $p(\vk)\!\rightarrow\!0$.
(iii)~For the noise-bias correction of \citet{Anacal_Li2025}, an independent
pure-noise realization with the same power spectrum as the image noise is rotated
by $90^\circ$, deconvolved with the correspondingly rotated PSF $p^{90}(\vk)$, and
reconvolved with the same $h(\vk)$ before being added to the image. The $90^\circ$
rotation makes the added noise statistically independent of the original noise
while preserving its second-order statistics, so that the ensemble-averaged shear
response is left unchanged and the noise-induced bias is canceled analytically.
The specific filter scales adopted in this work are quoted in
Section~\ref{subsec:method_fit}, and we refer the reader to
\citet{Anacal_Li2023, Anacal_Li2025} for demonstrations that the recovered shear
is insensitive to this choice, as long as the reconvolution Gaussian is broader
than the original PSF so that the deconvolution remains well conditioned.}

One limitation of this noise bias correction method is the need to artificially
double the image noise prior to detection and measurement. Notably, a recent
study by \citet{deepfield_metacal2023} proposes leveraging deep field images
from the same survey to mitigate noise effects in such correction schemes. We
plan to incorporate the methodology of \citet{deepfield_metacal2023} into our
analytical shear estimation framework in future work.

\subsection{Analytical Pixel Shear Response}
\label{subsec:method_pixres}

\citet{Anacal_Li2023} investigated the impact of weak-lensing shear distortions
on pixel values that have been smoothed by a Gaussian filter following PSF
deconvolution. Their analysis interprets these smoothed pixel values as
projections of the signal onto a set of pixel basis functions. The shear
response of each pixel values is derived using the analytical shear response of
the pixel basis functions. Building upon this framework, \citet{Anacal_Li2025}
derived analytical expressions for the shear response of the renoised image.

In this work, we present the shear response of the smoothed, renoised image
directly and refer the reader to \citet{Anacal_Li2023} and
\citet{Anacal_Li2025} for detailed derivations:
\begin{equation}
\label{eq:quintuple_image_deriv}
\begin{split}
    \frac{\partial f_{h}(\vx)}{\partial \gamma_1}
    &= G_1(\vx) + x_1 J_1(\vx) - x_2 J_2(\vx)\,, \\
    \frac{\partial f_{h}(\vx)}{\partial \gamma_2}
    &= G_2(\vx) + x_2 J_1(\vx) + x_1 J_2(\vx)\,,
\end{split}
\end{equation}
where the auxiliary smoothed images $G_1$, $G_2$, $J_1$, and $J_2$ are defined
as
\begin{equation}
\begin{split}
G_1(\vx) &=  \mathcal{F}^{-1}\left[
    \left(
        \frac{f_{p}(\vk)}{p(\vk)} - \frac{n(\vk)}{p^{90}(\vk)}
    \right) h(\vk) \;
    (k_1^2 - k_2^2)
\right]\,, \\
G_2(\vx) &= \mathcal{F}^{-1}\left[
    \left(
        \frac{f_{p}(\vk)}{p(\vk)} - \frac{n(\vk)}{p^{90}(\vk)}
    \right) h(\vk) \;
    (2\,k_1 k_2)
\right]\,, \\
J_1(\vx) &= \rmi\; \mathcal{F}^{-1}\left[
    \left(
        \frac{f_{p}(\vk)}{p(\vk)} - \frac{n(\vk)}{p^{90}(\vk)}
    \right) h(\vk) \;
    k_1
\right]\,, \\
J_2(\vx) &= \rmi\; \mathcal{F}^{-1}\left[
    \left(
        \frac{f_{p}(\vk)}{p(\vk)} - \frac{n(\vk)}{p^{90}(\vk)}
    \right) h(\vk) \;
    k_2
\right]\,,
\end{split}
\end{equation}
and $\mathcal{F}^{-1}(\bullet)$ denotes the inverse Fourier transform. The
quantities $G_1$ and $G_2$ characterize how the Gaussian PSF profile changes in
response to shear distortions along the $\gamma_1$ and $\gamma_2$ directions,
respectively. The combinations $x_1 J_1 - x_2 J_2$ and $x_2 J_1 + x_1 J_2$
describe how the relative positions of neighboring pixels are altered due to
the applied shear distortion. $G_1$, $G_2$, $J_1$, and $J_2$ will be used to
define quintuple numbers, a commutative ring structure, to automatically
propagate shear response to any observables measured from the smoothed,
renoised image.

\begin{figure*}
    \centering
    \includegraphics[width=.92\textwidth]{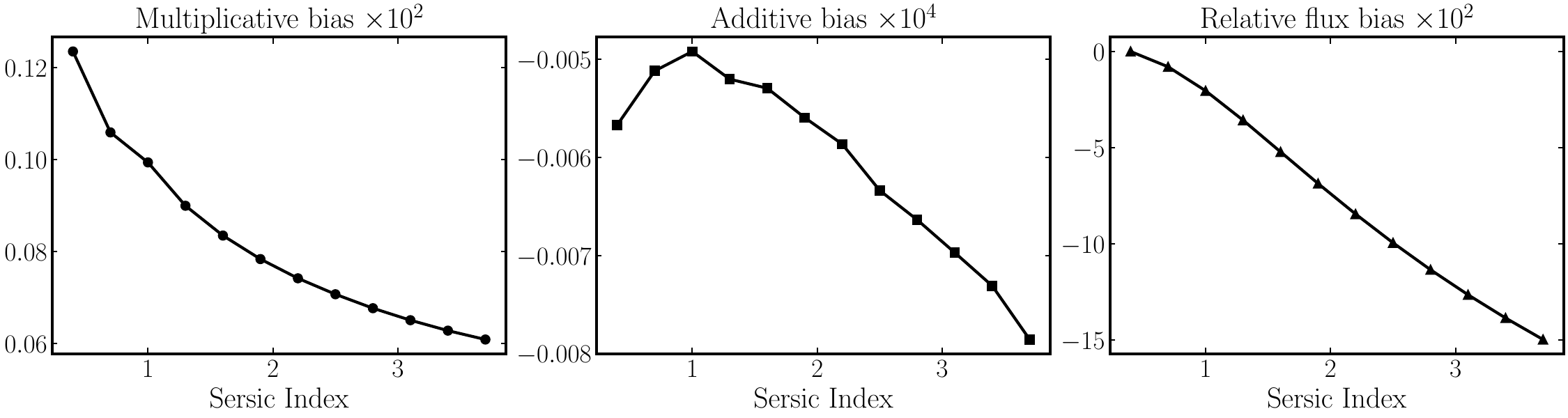}
\caption{
    The left and middle panels display the multiplicative ($m_1$) and additive
    ($c_1$) shear estimation biases, respectively, for galaxies with varying
    Sersic indices. The right panel shows the relative flux bias, defined as
    the ratio of measured to true flux minus one. Despite using a Gaussian
    model to fit galaxies with non-Gaussian morphologies---leading to a
    15\% flux bias---the shear estimation biases remain below the 0.2\%
    level relative to the input shear distortion ($\gamma_1 = 0.02$).
}\label{fig:model_bias}
\end{figure*}

We note that the shear response derived above assumes the origin of the shear
distortion is at the coordinate $(0, 0)$. If instead we wish to evaluate the
shear response relative to a different reference point, $(x_1^\text{ref},
x_2^\text{ref})$, the expressions for the pixel-wise shear response become:
\begin{equation}
\begin{split}
    \frac{\partial f_{h}(\vx)}{\partial \gamma_1}
    &= G_1(\vx) + (x_1 - x_1^\text{ref}) J_1(\vx) \\
    &- (x_2 - x_2^\text{ref}) J_2(\vx)\,, \\
    \frac{\partial f_{h}(\vx)}{\partial \gamma_2}
    &= G_2(\vx) + (x_2 - x_2^\text{ref}) J_1(\vx)\\
    &+ (x_1 - x_1^\text{ref}) J_2(\vx)\,.
\end{split}
\end{equation}
In Fig.~\ref{fig:pixel_res}, we present examples of the filtered images of a
noiseless galaxy that is used to determine its pixel shear response. The
reference origin point of the shear distortion is set to the center of the
image.

\subsection{Response Propagation with Quintuple Number}
\label{subsec:method_qnumber}

\begin{figure*}
    \centering
    \includegraphics[width=.92\textwidth]{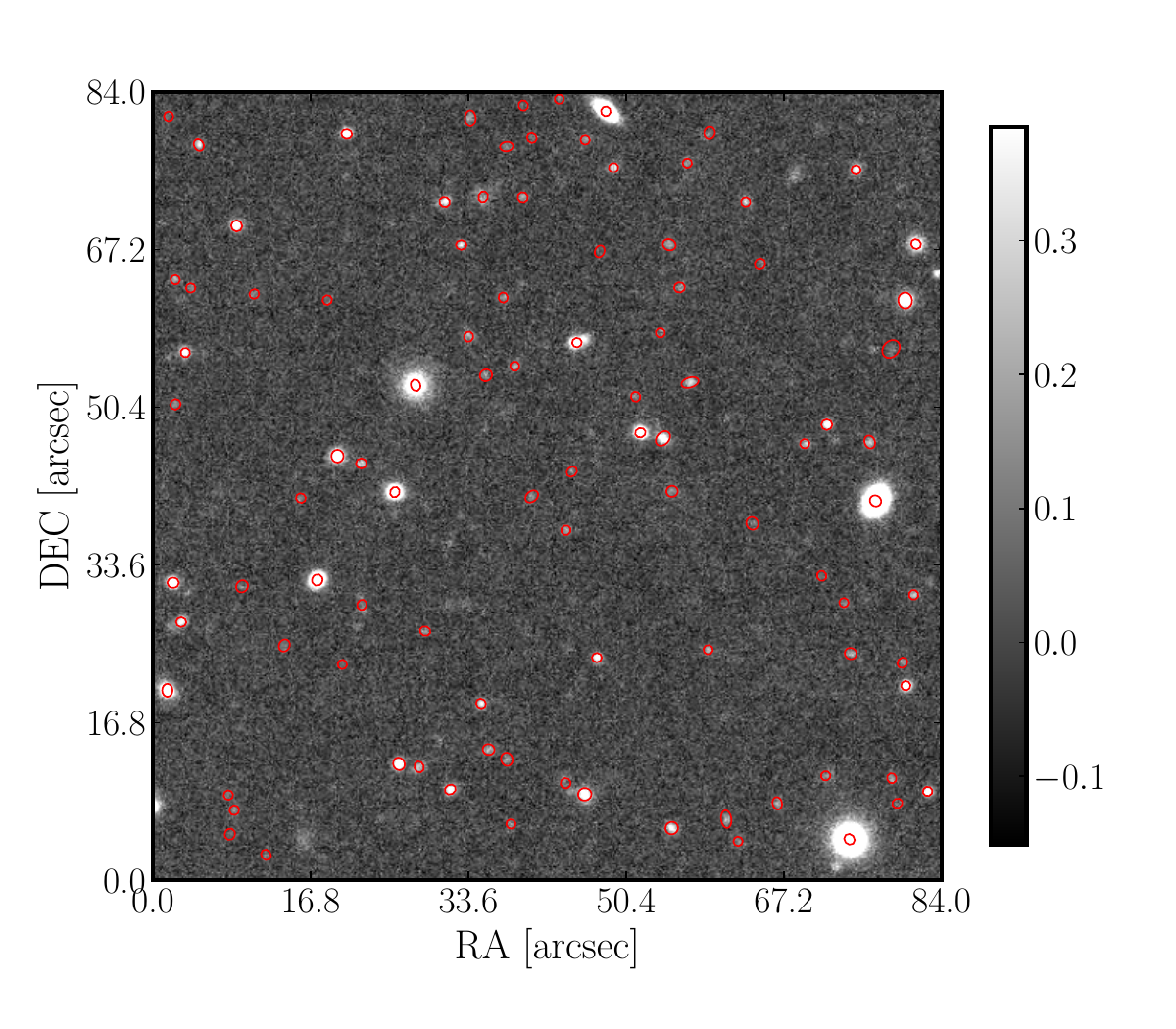}
\caption{
    We present a 2 deg$^2$ image simulation used to evaluate the accuracy of
    our shear estimator. Details of the simulation setup are provided in
    Section~\ref{sec:sim_sim}. The red ellipses indicate the 2$\sigma$ contours
    from Gaussian fits to the detected sources.
}\label{fig:sim_image}
\end{figure*}

With the pixel shear response derived in \citet{Anacal_Li2023} and presented in
equation~\eqref{eq:quintuple_image_deriv}, we can compute the shear response of
any observable measured from the smoothed image by applying the chain rule of
derivatives. In \citet{Anacal_Li2023}, this pixel-level shear response was used
to derive the shear response of \FPFS{} shapes \citep{FPFS_Li2018,
FPFS_Li2022}---a fixed-kernel shape estimator based on Shapelets
\citep{shapeletsI_Refregier2003, polar_shapelets_Massey2005}. Compared to
fixed-kernel estimators, model-fitting methods have higher signal-to-noise
ratio; however, they are significantly more complex, often requiring many
iterative steps. In principle, the shear response of a model-fitting estimator
can also be computed via the chain rule, provided all functions involved are
differentiable. However, manually tracing this response through the iterative
fitting process is generally intractable due to its complexity. To address this
challenge, we introduce quintuple numbers, an extension of dual numbers
\citep{autodiff} used in automatic differentiation, which allows us to
propagate shear response automatically through the model-fitting procedure.

We generalize the concept of infinitesimal $\epsilon$ in dual number
\citep{autodiff} by introducing four distinct infinitesimals, \(\epsilon_1,
\epsilon_2, \epsilon_3, \epsilon_4\), each of which squares to zero. Typically,
we also require that products of \emph{different} infinitesimals vanish as
well. More formally, we impose:
\begin{equation}
  \epsilon_i \,\epsilon_j = 0 \quad \text{for all } i, j\,.
\end{equation}
We then define a quintuple number:
\begin{equation}
  q = q_0 \;+\; q_1\,\epsilon_1 \;+\; q_2\,\epsilon_2 \;+\; q_3\,\epsilon_3
        \;+\; q_4\,\epsilon_4.
\end{equation}
Additionally, it can be expressed as a five-dimensional vector $(q_0, q_1, q_2,
q_3, q_4)$. Because all products \(\epsilon_i \epsilon_j\) vanish, this
five-term expression captures every possible term in the quintuple number.

Below, we give out the basic operation rules between two dual numbers $q^{(1)}
= \left(q_0^{(1)}, q_1^{(1)}, q_2^{(1)}, q_3^{(1)}, q_4^{(1)}\right)$ and
$q^{(2)}
= \left(q_0^{(2)}, q_1^{(2)}, q_2^{(2)}, q_3^{(2)}, q_4^{(2)}\right)$.

\textbf{Addition:}
Addition of two quintuple numbers is performed component-wise:
\begin{equation}
\begin{split}
q^{(1)} \,+\, q^{(2)} &\,=\, (q_0^{(1)} \,+\, q_0^{(2)})\\
    &\,+\,(q_1^{(1)} \,+\, q_1^{(2)})\,\epsilon_1
    \,+\,(q_2^{(1)} \,+\, q_2^{(2)})\,\epsilon_2\\
    &\,+\,(q_3^{(1)} \,+\, q_3^{(2)})\,\epsilon_3
    \,+\,(q_4^{(1)} \,+\, q_4^{(2)})\,\epsilon_4\,.
\end{split}
\end{equation}

\textbf{Multiplication:}
Multiplication of two quintuple numbers uses the distributive law and the fact
that any product \(\epsilon_i \epsilon_j\) is zero:
\begin{equation}
\begin{split}
q^{(1)} \,\times\, q^{(2)} &= (q_0^{(1)} \, q_0^{(2)})\\
    &+(q_0^{(2)}\, q_1^{(1)} \,+\, q_0^{(1)} \, q_1^{(2)})\,\epsilon_1\\
    &+(q_0^{(2)}\, q_2^{(1)} \,+\, q_0^{(1)} \, q_2^{(2)})\,\epsilon_2\\
    &+(q_0^{(2)}\, q_3^{(1)} \,+\, q_0^{(1)} \, q_3^{(2)})\,\epsilon_3\\
    &+(q_0^{(2)}\, q_4^{(1)} \,+\, q_0^{(1)} \, q_4^{(2)})\,\epsilon_4\,.
\end{split}
\end{equation}

\textbf{Division:}
For division by a non-infinitesimal (i.e.\ when the real part of the
denominator is nonzero), we keep only the first-order terms in each
\(\epsilon_i\):
\begin{equation}
\begin{split}
    \frac{q^{(1)}}{q^{(2)}} &\,=\, \frac{q_0^{(1)}}{q_0^{(2)}} \\
    &\,+\,\left(q_1^{(1)}\,q_0^{(2)} - q_1^{(2)}\,q_0^{(1)} \right)
    \left(q_0^{(2)}\right)^{-2}\epsilon_1 \\
    &\,+\,\left( q_2^{(1)}\,q_0^{(2)} - q_2^{(2)}\,q_0^{(1)} \right)
    \left(q_0^{(2)}\right)^{-2}\epsilon_2 \\
    &\,+\,\left( q_3^{(1)}\,q_0^{(2)} - q_3^{(2)}\,q_0^{(1)} \right)
    \left(q_0^{(2)}\right)^{-2}\epsilon_3 \\
    &\,+\,\left( q_4^{(1)}\,q_0^{(2)} - q_4^{(2)}\,q_0^{(1)} \right)
    \left(q_0^{(2)}\right)^{-2}\epsilon_4 \\
\end{split}
\end{equation}
Here again, we neglect any product of two \(\epsilon_i\)s.

Common analytic functions extend neatly to quintuple numbers by performing a
Taylor expansion in each infinitesimal direction and discarding second-order
(or higher) terms. For instance:

\textbf{Exponentiation:}
\begin{equation}
  \exp\bigl( q \bigr) \;=\; e^{q_0}
\Bigl(
    1 + q_1\,\epsilon_1 + q_2\,\epsilon_2 + q_3\,\epsilon_3 + q_4\,\epsilon_4
\Bigr)\,.
\end{equation}

\textbf{Power Functions:}
For \(n \in \mathbb{R}\), using the binomial expansion and discarding
second-order (or higher) terms:
\begin{equation}
q^n =\, q_0^n \,+\, n \, q_0 ^{\,n-1} \,
    \bigl(q_1\,\epsilon_1 + q_2\,\epsilon_4
    + q_3\,\epsilon_4 + q_4\,\epsilon_4\bigr).
\end{equation}

This quintuple system offers a straightforward and systematic method for
computing derivatives with respect to shear in the model-fitting operation,
eliminating the need for symbolic differentiation or finite differences. By
performing arithmetic directly on quintuple numbers, the partial derivatives
with respect to shear naturally propagate through algebraic operations and
function evaluations.

\xlrv{Because the four infinitesimal directions are fixed in advance, quintuple
numbers implement a highly specialized form of \emph{forward-mode} automatic
differentiation with a minimal, constant-size tangent space. This is what
distinguishes the approach from general-purpose differentiable-programming
frameworks such as \textsc{jax} \citep{jax2018}, which trace the sequence of
operations and just-in-time compile it, incurring tracing, compilation, and
dispatch overhead; reverse-mode differentiation through an iterative optimizer
additionally stores intermediate values whose memory footprint grows with the
number of operations. By contrast, each quintuple number is a lightweight
five-component object whose arithmetic and elementary functions (addition,
multiplication, division, exponentiation, powers, and trigonometric functions)
are evaluated in closed form directly in compiled code. The memory cost per
number is therefore constant, independently of the length or complexity of the
model-fitting computation. Furthermore, the fitting objective and its moments
are accumulated by summing over the pixels one at a time, so that each pixel's
quintuple contribution is released as soon as it has been added to the running
total and no per-pixel or per-operation derivative history is retained; the
peak memory therefore does not grow with the number of arithmetic operations or
optimizer iterations. Because these overloaded operations form a sufficient set
for the fitting objective and its iterative optimizer, the standard
model-fitting code runs essentially unchanged on quintuple-valued images, and,
since the two shear directions are seeded once in the input image, the shear
responses of all fitted parameters emerge from the same forward computation
that yields the parameters themselves---with no separate reverse pass or
finite-difference re-evaluation---at a cost of well under one millisecond per
galaxy.}

Concretely, we replace all input images in the model-fitting code with images
of quintuple numbers, denoted by
\begin{equation}
\mathcal{F}(\vx)= \left(
    f_h(\vx),\; \frac{\partial f_h (\vx)}{\partial \gamma_1},\;
    \frac{\partial f_h (\vx)}{\partial \gamma_2},\;
    J_1(\vx),\; J_2(\vx)
\right)\,.
\end{equation}
where $f_h$ is the smoothed renoised image, $\frac{\partial f_h}{\partial
\gamma_1}$, $\frac{\partial f_h}{\partial \gamma_2}$, $J_1$ and $J_2$ are
defined in equation~\eqref{eq:quintuple_image_deriv} measured from the image.
Additionally, we replace all arithmetic operations in the model-fitting code
with those defined for quintuple numbers, as summarized in this subsection. As
a result, the output of the model-fitting code is also a quintuple number. For
instance, we denote the fitted ellipticity by $\mathcal{E}_1$:
\begin{equation}
    \mathcal{E}_1 = \left(
    e_1,\; \frac{\partial e_1}{\partial \gamma_1},\;
    \frac{\partial e_1}{\partial \gamma_2},\;
    J^{e_1}_1,\; J^{e_1}_2
\right)\,.
\end{equation}
Thus, the fitted ellipticity and its shear responses, $\left(\frac{\partial
e_1}{\partial \gamma_1}, \frac{\partial e_1}{\partial \gamma_2}\right)$ are
automatically included in the quintuple output. The last two dimensions in the
quintuple number output $\mathcal{E}_1$ can be used to change the reference
point that of the shear distortion.

\xlrv{The choice of exactly five components is dictated by the
structure of the shear-response problem. Besides the value of the quantity
itself, we require its two first-order derivatives with respect to $\gamma_1$
and $\gamma_2$, together with the two spatial-gradient terms $J_1$ and $J_2$
that encode how the shear response transforms when the reference point of the
distortion is displaced---for example, from the image origin to the centroid of
a detected object (see equation~\eqref{eq:quintuple_image_deriv}). No further
components are needed, because every higher-order product of the infinitesimals
vanishes by construction, so the linear shear response is captured exactly.}

Finally, the shear can be estimated as
\begin{equation}
\begin{split}
    \hat{\gamma}_1 &= \langle e_1 \rangle \bigg/
        \left\langle \frac{\partial e_1}{\partial \gamma_1} \right\rangle\,,\\
    \hat{\gamma}_2 &= \langle e_2 \rangle \bigg/
        \left\langle \frac{\partial e_2}{\partial \gamma_2} \right\rangle\,.
\end{split}
\end{equation}
Even though we only use the first-order shear response, the estimated shear is
accurate to the second order of shear since the second order of shear
$\gamma_{1,2}^2$ has different spin number to the galaxy shape
$\bar{\gamma}_{1,2}$. Therefore, we have
\begin{equation}
\label{eq:shear_estimator}
    \hat{\gamma}_{1,2} = \gamma_{1,2} + \mathcal{O}(\gamma^3)\,.
\end{equation}

\subsection{Model Fitting}
\label{subsec:method_fit}

In this paper, we implement a simple Gaussian model to fit galaxy shapes,
sizes, and fluxes, and demonstrate that our quintuple number system enables
accurate propagation of the shear response through the model-fitting process.
More sophisticated multi-component fitting approaches (e.g.,
\citet{lensfit2024}) will be explored in a future publication.

We model the object as an elliptical 2D Gaussian with major axis length $a_1$,
minor axis length $a_2$, and orientation angle $\theta$. The covariance matrix
of the original Gaussian is:
\begin{equation}
    \Sigma = Q
    \begin{pmatrix}
    a_1^2 & 0 \\
    0 & a_2^2
    \end{pmatrix}
    Q^\top,
    \quad \text{where} \quad
    Q =
    \begin{pmatrix}
    \cos\theta & -\sin\theta \\
    \sin\theta & \cos\theta
    \end{pmatrix}
\end{equation}
s the rotation matrix corresponding to an angle $\theta$\,.

Since the galaxy model is fit to the reconvolved image (see
equation~\eqref{eq:quintuple_image_value}), the Gaussian model must be
convolved with the point spread function (PSF). We convolve this profile with
the isotropic Gaussian PSF of width $\sigma_h$. The covariance matrix of the
PSF is:
\begin{equation}
    \Sigma_{\text{PSF}} = \sigma_h^2 I =
    \begin{pmatrix}
    \sigma_h^2 & 0 \\
    0 & \sigma_h^2\,.
    \end{pmatrix}
\end{equation}
The resulting convolved Gaussian has a covariance matrix given by the sum:
\begin{equation}
    \Sigma_\text{o} = \Sigma + \Sigma_{\text{PSF}}\,.
\end{equation}
The PSF-convolved Gaussian is expressed as:
\begin{equation}
    M(x, y) = \frac{F}{2\pi \sqrt{\det(\Sigma_\text{o})}}
    \exp\left( -\frac{1}{2}
    \vx'^T \Sigma_\text{o}^{-1} \vx'
    \right)\,,
\end{equation}
where $\vx' = \vx - \vx_c$, and $\vx_c$ is the center of the Gaussian. The
galaxy shape is defined as
\begin{equation}
\begin{split}
    e_1 &= \frac{a_1^2 - a_2^2}{a_1^2 + a_2^2} \cos{2\theta}, \\
    e_2 &= \frac{a_1^2 - a_2^2}{a_1^2 + a_2^2} \sin{2\theta}\,.
\end{split}
\end{equation}

We implement this Gaussian model-fitting algorithm with our quintuple number
system and the shear estimation procedure consists of the following steps:
\begin{enumerate}
    \item \textbf{Quintuple Number Transformation:} Convert the exposure image
        into a \emph{quintuple number image} following
        equations~\eqref{eq:quintuple_image_value} and
        \eqref{eq:quintuple_image_deriv}, where each pixel carries not only its
        intensity but also its shear response derivatives, enabling analytical
        propagation of shear effects.

    \item \textbf{Galaxy Detection:} Identify galaxy candidates directly from
        the quintuple number image using the differentiable detection algorithm
        described in \citet{Anacal_Li2025}. This step ensures that the detected
        objects already encode their response to shear.

    \item \textbf{Gaussian Model Fitting:} For each detected object, perform
        independent Gaussian model fitting to estimate galaxy shape and flux
        parameters. The shear response is analytically propagated through the
        model-fitting process using the quintuple number formalism.
\end{enumerate}

We simulate a set of intrinsically isotropic galaxies modeled with Sersic
profiles \citep{Sersic1963} of varying indices, representing a range of
realistic galaxy morphologies. Each galaxy is sheared by a small input
distortion of $\gamma_1 = 0.02$. A simple Gaussian model-fitting algorithm is
then applied to estimate both the galaxy shape and flux from the observed
image. All galaxies have a half-light radius of $0\farcs20$ and are convolved
with a Moffat point spread function (PSF) \citep{Moffat1969}, defined as
\begin{equation}
\label{eq:moffat_PSF}
    p(\vx) = \left[1 + c\left(\frac{|\vx|}{r_\mathrm{P}}\right)^2\right]^{-2.5},
\end{equation}
with parameters $r_\mathrm{P}$ and $c$ chosen such that the full width at half
maximum (FWHM) is $0\farcs60$, consistent with the median seeing of the HSC
survey \citep{HSC3_catalog}. Before measuring shapes and fluxes, we circularize
the PSF by transforming it into an isotropic Gaussian with FWHM $0\farcs70$.
The estimated shapes are then converted to shear using
equation~\eqref{eq:shear_estimator}. To quantify the accuracy of shear
recovery, we evaluate the standard multiplicative ($m_{1,2}$) and additive
($c_{1,2}$) bias parameters \citep{shearSys_Huterer2006, Heymans2006}, defined
through the relation:
\begin{equation}
\label{eq:shear_biases}
\hat{\gamma}_{1,2} = (1 + m_{1,2})\,\gamma_{1,2} + c_{1,2}.
\end{equation}
The results are shown in Fig.~\ref{fig:model_bias}. Despite fitting
non-Gaussian galaxies with a Gaussian model---which introduces a flux bias of
up to 15\%---the resulting shear estimation biases remain below the 0.2\% level
relative to the input shear.

\section{Test On Image Simulation}
\label{sec:sim}

\subsection{Simulation}
\label{sec:sim_sim}

For our image simulations, we use the
\texttt{descwl-shear-sims}\footnote{
    \url{https://github.com/LSSTDESC/descwl-shear-sims}
}
package, described in \citet{metaDet_LSST2023}. This tool generates realistic
synthetic galaxy images using the \galsim{} framework \citep{GalSim}. The
outputs include background-subtracted, photometrically calibrated images, along
with corresponding noise variance maps. Each simulation also includes a world
coordinate system (WCS) and a point spread function (PSF) model.

To isolate the performance of the shear estimator under ideal conditions, we
deliberately exclude sources of systematic error such as PSF modeling
inaccuracies, noise misestimation, image warping artifacts, coaddition errors,
and astrometric or photometric miscalibration. These effects are omitted to
ensure that our tests reflect performance under well-controlled, coadded image
conditions. While our current focus is on validation under these simplified
circumstances, a complete shear calibration for real data will eventually
require accounting for such imperfections.

All simulations include blended galaxy systems. A uniform shear is applied
across each subfield, such that all galaxies within a given image experience
the same lensing distortion. Both galaxy shapes and positions are transformed
according to the applied shear, which takes values of $\gamma_1 = \pm 0.02$ in
our tests. Each simulation scenario comprises 5,000 subfields, each covering an
area of $0.036,\mathrm{deg}^2$.

To suppress shape noise, we adopt the ring test method \citep{galsim_STEP2},
pairing each subfield with a counterpart rotated by $90^\circ$. Additionally,
we apply the shape noise cancellation technique proposed by
\citet{preciseSim_Pujol2019}, which further reduces uncertainty in shear
measurements arising from intrinsic galaxy shape dispersion and measurement
noise. Simulations assume a circular Moffat PSF as defined in
equation~\eqref{eq:moffat_PSF} with HSC seeing size FWHM equals $0\farcs60$
\citep{HSC3_catalog}, and we refer readers to \citet{Anacal_Li2025} for tests
with anisotropic PSFs.

To approximate the effect of residual Poisson noise due to high sky background,
Gaussian noise is added after background subtraction. Noise levels are
estimated using the \texttt{WeakLensingDeblending}\footnote{
    \url{https://github.com/LSSTDESC/WeakLensingDeblending}} package
\citep{weaklensDeblend}, which is configured according to LSST filter
characteristics. After adding all relevant features and noise, images are
normalized to a photometric zero point of 27.

Unless otherwise specified, we simulate coadded images in the ``$i$'' band
using the HSC year 3 depth \citep{HSC3_catalog} as our standard configuration.
The input galaxies are modeled with \texttt{WeakLensingDeblending} and include
bulge, disk, and AGN components. Morphologies are consistent across bands,
though flux and ellipticity can vary between components. The AGN is represented
as a central point source. The raw source density is 240 galaxies per square
arcminute, with an effective $i$-band AB magnitude limit of 27. Galaxies are
uniformly distributed without accounting for clustering.

\xlrv{We emphasize that these input galaxies are rendered as
bulge$+$disk$+$AGN systems drawn from a catalog whose joint
magnitude--size--color distributions are calibrated to reproduce real
observations, and the catalog extends well below the magnitude cut applied in
the analysis. The simulated galaxies therefore span a broad, realistic range of
sizes rather than a single fixed size. The uniform shear applied to each
subfield is a constant (non-cosmological) distortion shared by all galaxies
irrespective of their redshift, as is standard for shear-calibration tests; it
is deliberately \emph{not} a redshift-dependent cosmological shear field. The
same-redshift condition therefore enters only through the treatment of the
lensing shear distortion---so that all sources, including blended ones, share a
single shear---as discussed in the introduction, and not through the size or
redshift distribution of the sources themselves.}

To reduce statistical noise in bias estimation, we make use of both $90^\circ$
rotated galaxy pairs \citep{galsim_STEP2} and the shape noise suppression
method from \citet{preciseSim_Pujol2019}. An example of the simulation is shown
in Fig.~\ref{fig:sim_image}.

\subsection{Result}
\label{sec:sim_res}

\begin{figure}
\centering
    \includegraphics[width=.45\textwidth]{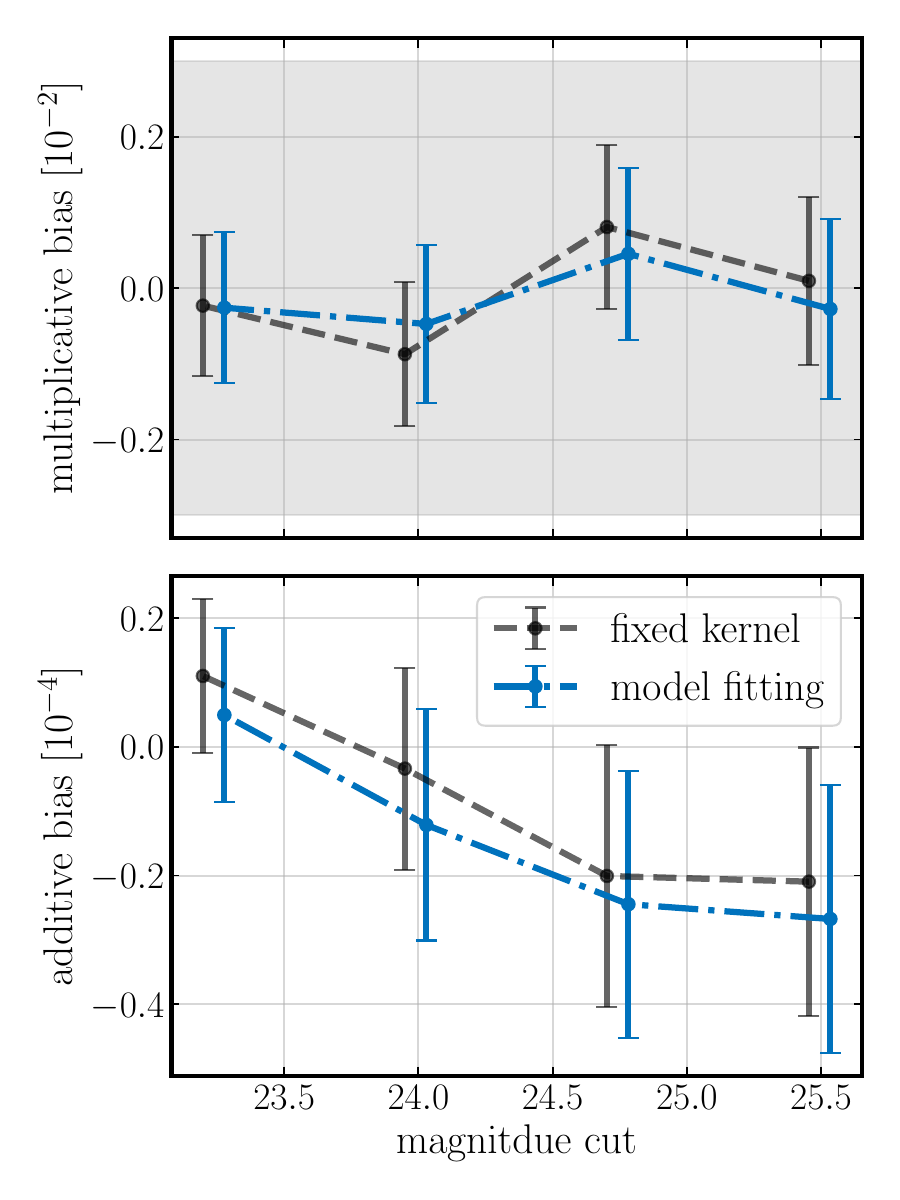}
\caption{
    The multiplicative (upper panel) and additive bias (lower panel) as a
    function of magnitude cut. The black lines show the result for fixed-kernel
    shape estimator (\FPFS{}) and the blue lines are for model-fitting code
    introduced in Section~\ref{subsec:method_fit}. The shaded region shows the
    requirement on the control of multiplicative bias in LSST Dark Energy
    Science Collaboration \citep{LSSTRequirement2018}.
}\label{fig:shear_bias}
\end{figure}

We assess the accuracy of our Gaussian model fitting algorithm introduced in
Section~\ref{subsec:method_fit}, which has been fully implemented using the
quintuple number system introduced in Section~\ref{subsec:method_qnumber}. This
quintuple number system enables the analytical propagation of shear responses
through the model-fitting process by automatically tracking first-order
derivatives with respect to shear distortions. The implementation allows us to
evaluate how the measured galaxy shape, size, and flux respond to small changes
in gravitational shear without relying on finite-difference approximations.

In addition to testing model fitting, we also reimplemented our Fixed kernel
shear estimator (\FPFS{}; \citet{FPFS_Li2018, FPFS_Li2022}) estimator within
the same quintuple number framework. The \FPFS{} method estimates shear using
second-order \citep{FPFS_Li2018} and fourth-order \citep{FPFS4th} Shapelets
\citep{shapeletsI_Refregier2003, polar_shapelets_Massey2005} weight function
applied to image pixels and it is optimized in \citet{FPFS_Li2023b}. By
integrating it into the quintuple system, we can now derive the shear response
of \FPFS{} estimators analytically and automatically, streamlining the
calibration process while improving numerical robustness.

The figure illustrates the performance of both methods. The top panel shows the
multiplicative bias ($m_1$ in equation~\eqref{eq:shear_biases}) in the
recovered shear estimates, while the bottom panel shows the corresponding
additive bias ($c_1$ in equation~\eqref{eq:shear_biases}). These biases are
plotted as a function of magnitude cuts, where the magnitudes are computed from
the fluxes returned by the model fitting.

For both the Gaussian model fitting and the \FPFS{} method, the multiplicative
biases remain well within the requirements set by the LSST Dark Energy Science
Collaboration \citep{LSSTRequirement2018} for controlling shear calibration
bias. Additionally, the additive biases are statistically consistent with zero,
indicating that neither method introduces significant directional systematics
into the shear measurements.

\section{SUMMARY AND OUTLOOK}
\label{sec:summary}

In this paper, we have extended the \anacal{} formalism to enable analytical
shear response calculations for galaxy model fitting. By introducing the
\emph{quintuple number system}, a commutative ring inspired by dual numbers
from automatic differentiation, we developed a method that can automatically
propagate shear response information from image pixels to model-fit parameters,
such as galaxy flux, shape, and size. This advancement allows us to
analytically evaluate how small shear distortions impact the outcomes of model
fitting, removing the need for finite differencing or image perturbation
techniques.

We implemented a Gaussian model-fitting pipeline fully integrated with the
quintuple number system, allowing us to derive shear responses alongside
best-fit parameters in a single, automated process. Additionally, we
reimplemented our \FPFS{} estimator within the same algebraic system. This
enabled the \FPFS{} method, which uses a fixed Gaussian kernel for shear
estimation, to inherit automatic shear response derivation, thus benefiting
from the same analytical framework.

We validated both the Gaussian model-fitting estimator and the \FPFS{}
estimator using realistic image simulations covering 180~deg$^2$ of sky with
observing conditions modeled after the Hyper Suprime-Cam survey. The upper
panel of our result figure shows the multiplicative bias for both methods,
while the lower panel displays the corresponding additive bias. Both estimators
achieve multiplicative biases well within the LSST shear bias requirements
(i.e., below $3 \times 10^{-3}$). The additive biases are consistent with zero,
indicating that neither method introduces significant residual shear
systematics.

Our results confirm that the quintuple number framework provides a robust,
fully analytical, and computationally efficient foundation for future shear
estimation pipelines. By bridging the pixel-level shear response to high-level
galaxy parameter inference, this framework offers a promising path toward
high-precision, survey-ready shear calibration.

In future work, we plan to extend the quintuple number framework to support
more sophisticated model-fitting algorithms, including multi-component and
Sérsic-based models that more accurately capture the diversity of galaxy
morphologies (see e.g. \citet{lensfit2024}). \xlrv{To improve the
\emph{precision} of shear estimation, we further plan to combine a convolutional
neural network for galaxy shape estimation with the quintuple number system for
the analytical shear calibration of the CNN-based shapes, building on the
approach of \citet{Lin2026, Lin2026b}.} Additionally, we aim to integrate a deblending module within
the quintuple number system, allowing analytical shear responses to be
propagated through the deblending process \citep{2024A&C....4900875S}.

\acknowledgments

This paper makes use of software developed for the Vera C.\ Rubin Observatory.
We thank the Vera C.\ Rubin Observatory for making their code available as free
software at http://dm.lsst.org.

We thank the maintainers of \numpy{} \citep{numpy_Harris2020}, \texttt{SciPy}
\citep{scipy_Virtanen2020}, \texttt{Matplotlib} \citep{matplotlib_Hunter2007},
\galsim{} \citep{GalSim} and \texttt{conda-forge} \citep{conda_forge} projects
for their excellent open-source software and software distribution systems.

We thank Erin Sheldon, Alan Zhou, Rachel Mandelbaum, Matthew Becker and Scott
Dodelson for useful discussions.

Xiangchong Li acknowledges support from the U.S. Department of Energy under
Contract No. DE-SC0012704 and from the Laboratory Directed Research and
Development (LDRD) Program at Brookhaven National Laboratory (Project No.
27992).

\section*{Data Availability}
The code used for this paper is publicly available on Github:
\url{https://github.com/mr-superonion/AnaCal/tree/v0.6.0}

\bibliographystyle{aasjournal}
\bibliography{citation}

\end{document}